**Authors Details:**

**-** Soheila Khajoui
- Saeid Dehyadegari
- Sayyed Abdolmajid Jalaee

**Affiliations**
- Department of Management, Faculty of Management & Economics, Shahid Bahonar University of Kerman, Kerman, Iran
- Department of Management, Faculty of Management & Economics, Shahid Bahonar University of Kerman, Kerman
- Department of Economics, Faculty of Management & Economics, Shahid Bahonar University of Kerman, Kerman, Iran

**Author's email address**

- s.khajoui@aem.uk.ac.ir

- dehyadegari@uk.ac.ir

- jalaee@uk.ac.ir

**Declaration of interests**
The authors declare that they have no known competing financial interests or personal relationships that could have appeared to influence the work reported in this paper.

# Forecasting Imports in OECD Member Countries and Iran by Using Neural Network Algorithms of LSTM

**Abstract**

Artificial Neural Networks (ANN) which are a branch of artificial intelligence, have shown their high value in lots of applications and are used as a suitable forecasting method. Therefore, this study aims at forecasting imports in OECD member selected countries and Iran for 20 seasons from 2021 to 2025 by means of ANN. Data related to the imports of such countries collected over 50 years from 1970 to 2019 from valid resources including World Bank. WTO, IFM and…the data turned into seasonal data to increase the number of collected data for better performance and high accuracy of the network by using Diz formula that there were totally 200 data related to imports. This study has used LSTM to analyze data in Pycharm. 75% of data considered as training data and 25% considered as test data and the results of the analysis were forecasted with 99% accuracy which revealed the validity and reliability of the output. Since the imports is consumption function and since the consumption is influenced during Covid-19 Pandemic, so it is time-consuming to correct and improve it to be influential on the imports, thus the imports in the years after Covid-19 Pandemic has had a fluctuating trend.

**Keywords**: Artificial intelligence, Import Forecasting, International Trade, LSTM neural network, modeling

**JEL Classification:** F17, B26, C45

1. **Introduction**

    International trade reinforces economic growth (Redmond and Nasir 2020). People traded from a very long time ago to fulfill the need of themselves and their society and obtain more profits. International trade is in fact exchanging goods among countries and trading maybe private or governmental which is done by the merchants of international trade and international law (Zinatian 2015, 18). Multinational countries which are active in Global Value Chain (GVC) are also active in international trade. Investment rules were made due to simultaneous increase of intermediate good trade, intra-company trade and Foreign Direct Investment (FDI) (Heid and Vozzo 2020). Since the countries are not able to produce all their needed goods and also since it is not economical to produce all goods by themselves, so they exchange goods with each other. Exports and imports are the instruments which measure the economic ability of each country. Imports also influences the efficiency of trading policies, recognizing macroeconomics patters, the process of production, growth and development (Spanlo and Ghanbari 2011). Import as one of the factors of production plays an important role in economic growth because for the continuation of production and the growth of its capacities, the import of capital and intermediate goods is vital in economy (Shafeei et al 2020). With the rapid development of artificial intelligence and computer technology, the business forecasting model is constantly updating and integrating. Neural network modeling is based on internal data mechanism which is useful in recognizing pattern, intelligent control, signal processing and other fields (Zhang and Lou 2021), so ANN is used for forecasting. Today's Artificial Intelligence

(AI) systems are information systems which logically act what they know. AI refers to information systems which logically act upon the existing information to solve the problems. Despite so many applications of AI on lots of industries, all AI systems may be explained by using a common input-processing model (Paschen et al. 2020). Neural networks based on the concept of imitation from human's brain has features like generalizing samples and unsupervised or without supervision training. Practically, they are nonlinear statistical data modeling or decision-making instruments which may be used to model complicated relationships between inputs and outputs without the need to explain fundamental phenomena (Stalidis et al. 2015). ANNs are so useful for achieving accurate models and forecasts with low computational costs by using the data of the existing sample (Lot et al. 2020). Machine learning and neural network is widely used in different fields including agriculture, trade, automatic transportation vehicles, medicine and even artificial creativities (Tsai et al. 2021). Machine learning and neural network are used for solving complicated problems and have a structure consisted of interlocking blocks which are mixed together (O. Nikitin et al. 2021). AI and neural network are used in trade to forecast market fluctuations and analyze customer behavior and also market segmentation (Mustak et al. 2020). Since imports has lots of fluctuations and so many factors influence imports trade, traditional algorithms have weak performance in achieving accurate forecast results. Also, since it is difficult to accurately forecast imports and it needs so much time, cost and human force, so ANN algorithms in forecasting imports may be an effective method and forecast with high accuracy. Therefore, this study aims at forecasting imports trend in OCED countries and Iran by means of LSTM neural network.

2. **Literature review**

Kaplan and Haenlein (2021) believe that when AI is with robotics and IoT, it begins entering into all society fields and appears in popular and commercial press such as imports about what is generally called Industry 4.0. Alam (2019) states that two ARIMA and neural network methods are used to forecast annual imports and exports of Saudi Arabia, ARIMA, a traditional method is not suitable for time series with much data and todays ANNs are widely used to forecast time series with high data volume. Nik Aein (2017) identified and prioritized effective factors on importers' decision to import clothing to Iran market. Nowadays, good imports play a prominent role in Iran foreign trade so that billion-dollar goods including consumables, investment and intermediate enter Iran through different customs sources. On the other hand, somewhat intense economic changes in near future such as Iran sanctions removal, makes it necessary to pay special attention to import foreign products; therefore, it is important to identify and prioritize effective factors on decision-makers to Iran market. Li Shen et al. (2021) studied effective multinational trade forecast by means of LSTM neural network and concluded that neural network methods have an accurate performance for forecasting. Yousefi et al (2013) studied crude oil request forecast in Iran by means of neural networks and ARIMAX model and stated that neural network enjoys more accuracy rather than ARIMAX model to forecast crude oil request. R.Luchko, Dzubanovska and Arzamasova (2021) studied exports and imports forecast by means of neural network and concluded that ANNs are the most successful models to forecast exports and imports. D. Urrutia, M. Abdul and

Atienza (2019) studied Philippine imports and exports forecast and concluded that neural network methods are the most suitable forecast methods.

Since so many studies have been conducted on imports forecast by different methods, the current study has been conducted by LSTM neural network with high accuracy and ability to recognize complicate equations among data including economic data to forecast imports in OCED selected countries and Iran consisted of three developed, developing and underdeveloped countries. Selecting this method and also mentioned countries group added to the importance and innovation of this plan and also the related literature.

3. **Model**

It is a forecast of modern mathematics and data science application. Over 100 past years, forecasting methods and models have been so improved. So that machine learning and AI methods have been developed as a powerful substituent for traditional methods for forecasting (Koliadenko 2021). AI and IT are important phenomena for B2B (Business-to-Business). The current trends show that AI and IT help the difficulty reduction of compact information tasks (Keranen and Prior 2019). AI systems are able to effectively process much values of different data such as structured and non-structured and their ability to process non-structured data shows the value and advantage of this method which distinguishes it from other traditional methods (Carrillo 2020). Among the forecasting models appropriate with financial time series used in business are ANNs which act well in recognizing pattern in much values of data (like time series) (Vogl et al. 2022). Considering the formation of time series data with much volume, it seems necessary to create a LSTM to optimize forecasting to be able to recognize more than 100 hundred step time dependencies and also recognize complicated interactions of financial and economic data (Siami-Namin 2018). Like Recurrent Neural Network (RNN), LSTM has recurrent connections, but unlike RNNs, LSTM has a unique feature which avoids problems which exist in traditional RNN training and scaling. Thus, effective results are made by this network which have made LSTM more popular (Greff et al. 2017).

Imports demand depends on exchange resources and incomes; therefore, imports demand function is as follows:

$$LM_t = b_0 + b_1 LY_t + b_2 LP_t + b_3 LF_t + b_4 LR_{t-1} + b_5 LM_{t-1} + u_t \qquad (1)$$

L is natural logarithm, $Y_t$ is national income, $P_t$ is imports relative prices, $F_t$ is exchange incomes and $R_{t-1}$ and $M_{t-1}$ are exchange reservoirs and imports of the past period (Moran 1989).

4. **Method**

This is a descriptive study and it is applied in purpose. The necessary data collected by library method and by using time series data. To forecast imports in Iran and OECD member countries such as the U.S, Canada, Germany, France, Japan, Turkey, South Korea, Portugal and Greece, the data related to imports extracted from international valid websites including World Bank, WTO, IFM and… over 1970 to 2019. Since the increasing the number of data causes neural network accuracy and performance, such annual data are turned into seasonal one by using Diz method. So, the volume of data used for each country was 200 data. Also, LSTM has been used for forecasting.

### 4.1. LSTM Neural Network

ANNs are used in many fields including engineering and sciences that traditional methods are not able to model and recognize the problem. ANNs are inspired by human' brain nervous system and are used as a good and accurate method for forecasting (Tuan Hoang et al. 2021). LSTM is a type of ANNs. LSTM is a special kind of RNNs with unique features. This network can remember data trend and sequence by using recurrent connections (Siami-Namin 2018). Also, LSTM has been made up of one memory cell which saves up-to-date information through input gates, forget gates and output gate (Koliadenko 2021). (figure 1). This network removes problems in traditional RNN training and scaling (Fan et al. 2021).

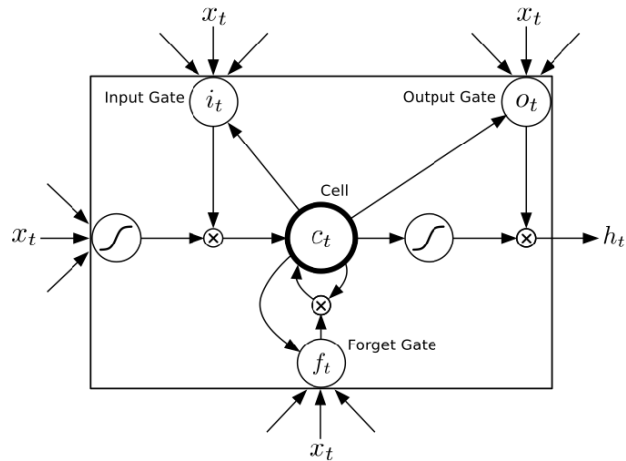

Figure1: Internal structure of LSTM neural network (Koliadenko 2021).

Mathematical description of information flow in LSTM neural network is stated as follows (⊕ and ⊗ are addition and multiplication). The arrows show the direction of information flow.

The first layer of memory gate shows removed unnecessary information to cellular status and is stated as follows:

$$f_t = \sigma(w_f \times x_t + U_f \times h_{t-1} + b_f) \qquad (2)$$

Where $F_t$ is forgetting threshold at time t, σ is sigmoid activation function, $W_f$ and $U_f$ are weights, $x_t$ is input value, $h_{t-1}$ is output at time t-1 and $b_f$ is bias.

The second input gate determines which information has to be saved in the cellular status of the current input vector. The mathematical equation of this gate is as follows:

$$i_t = \sigma(w_i \times x_t + U_i \times h_{t-1} + b_i) \qquad (3)$$

$$\overline{c_t} = \sigma(w_c \times x_t + U_c \times h_{t-1} + b_c \qquad (4)$$

$i_t$ shows input threshold at time t, $c_t$ is new state, $w_i$, $U_i$, $w_c$ and $U_c$ are weights value and $b_i$ and $b_c$ are bias.

The following equation has to be used to up to date cell status at time t:

$$c_t = f_t \times c_{t-1} + i_t \times \overline{c_t} \qquad (5)$$

The third layer is created as output information at the current status. Its mathematical equation is as follows:

$$o_t = \sigma(w_o \times x_t + U_o \times h_{t-1} + b_o) \qquad (6)$$

$o_t$ is output threshold at time t, $w_o$ and $U_o$ are weights and $b_o$ are bias.

at last, cell output us determined by the following formula:

$$h_t = o_t \times \tanh c_t \qquad (7)$$

$h_t$ is cell output at time t, tanh is activator function agent, $c_t$ shows cell sates at time t. When data passes through these three gates, the effective information is output and the invalid information is forgotten (Fan et al. 2021).

5. **Results**

As mentioned before, the more input data volume applied to neural network, the higher the extracted output accuracy will be. Therefore, the data in this study is related to 50 years (from 1970 to 2019) which is related to imports in Iran and 9 OECD member countries including the U.S, Canada, Germany, France, Japan, Turkey, South Korea, Portugal and Greece and the data turned into seasonal data by means of Diz formula. Input data volume for each country turns into 200 data for each country. LSTM neural network was implemented by Pychram with Keras library function by two test data and train data segmentation methods. The data segmented into 25% test data and 75% train data. Then, LSTM network structure and architecture were implemented. This network is consisted of three neuron layers. Then the neural network is trained and training repeated for 200 times and the state with less MSE and MAE rather than other states considered as the most suitable state. Tables 1 and 2 show MSE and MAE results for train data and test data to forecast imports for each country based on the normalized data. Accordingly, table 2 shows data analysis results to forecast imports in the mentioned countries over 20 seasons (from 2021 to 2025) with 99% accuracy.

**Table 1-Forecasting error of LSTM neural network Train data**

Source: Author's Calculations

|  | MSE | MAE |
|---|---|---|
| US | 0.00026587 | 0.009558063 |
| Canada | 0.000573174 | 0.017034501 |
| Germany | 0.000610407 | 0.016112808 |
| France | 0.000567954 | 0.017604845 |
| Japan | 0.000913517 | 0.020788627 |
| Turkey | 0.000534323 | 0.013281038 |
| Korea | 0.000237155 | 0.009067117 |
| Portugal | 0.000306369 | 0.011394773 |
| Greece | 0.000185511 | 0.009552464 |
| Iran | 0.000538507 | 0.01618062 |

**Table 2-Forecasting error of LSTM neural network test data**

Source: Author's Calculations

|  | MSE | MAE |
|---|---|---|
| US | 0.000593822 | 0.010549653 |
| Canada | 0.000471171 | 0.016917394 |
| Germany | 0.000437126 | 0.015138935 |
| France | 0.000409339 | 0.016679849 |
| Japan | 0.000633532 | 0.01640135 |
| Turkey | 0.000314573 | 0.01217957 |
| Korea | 0.000636021 | 0.012340967 |
| Portugal | 0.000986045 | 0.01962471 |
| Greece | 0.000188108 | 0.008993208 |
| Iran | 0.000370002 | 0.013099987 |

**Table 3- Import forecast based on LSTM neural network (current US$)** (Source: Author's Calculations)

|  |  | US | Canada | Germany | France | Japan | Turkey | Korea | Portugal | Greece | Iran |
|---|---|---|---|---|---|---|---|---|---|---|---|
| 2021 | q1 | 3E+12 | 5.02E+11 | 1.62E+12 | 7.03E+11 | 7.48E+11 | 2.62E+11 | 5.94E+11 | 8.11E+10 | 7.36E+10 | 5.53E+10 |
|  | q2 | 3.03E+12 | 4.96E+11 | 1.65E+12 | 6.91E+11 | 7.29E+11 | 2.68E+11 | 6.04E+11 | 7.92E+10 | 7.36E+10 | 5.24E+10 |
|  | q3 | 3.06E+12 | 4.91E+11 | 1.68E+12 | 6.88E+11 | 7.13E+11 | 2.72E+11 | 6.2E+11 | 7.85E+10 | 7.4E+10 | 4.94E+10 |
|  | q4 | 3.08E+12 | 4.88E+11 | 1.7E+12 | 6.93E+11 | 6.99E+11 | 2.74E+11 | 6.37E+11 | 7.9E+10 | 7.43E+10 | 4.63E+10 |
| 2022 | q1 | 3.09E+12 | 4.86E+11 | 1.71E+12 | 7.01E+11 | 6.91E+11 | 2.76E+11 | 6.45E+11 | 8.05E+10 | 7.47E+10 | 4.34E+10 |
|  | q2 | 3.1E+12 | 4.84E+11 | 1.71E+12 | 7.12E+11 | 6.85E+11 | 2.76E+11 | 6.58E+11 | 8.25E+10 | 7.55E+10 | 4.07E+10 |
|  | q3 | 3.11E+12 | 4.83E+11 | 1.7E+12 | 7.23E+11 | 6.82E+11 | 2.76E+11 | 6.65E+11 | 8.47E+10 | 7.59E+10 | 3.81E+10 |
|  | q4 | 3.11E+12 | 4.82E+11 | 1.68E+12 | 7.33E+11 | 6.82E+11 | 2.75E+11 | 6.67E+11 | 8.68E+10 | 7.63E+10 | 3.55E+10 |
| 2023 | q1 | 3.1E+12 | 4.81E+11 | 1.66E+12 | 7.4E+11 | 6.83E+11 | 2.74E+11 | 6.66E+11 | 8.87E+10 | 7.66E+10 | 3.29E+10 |
|  | q2 | 3.1E+12 | 4.79E+11 | 1.64E+12 | 7.43E+11 | 6.85E+11 | 2.72E+11 | 6.63E+11 | 9.01E+10 | 7.66E+10 | 3.02E+10 |
|  | q3 | 3.09E+12 | 4.78E+11 | 1.63E+12 | 7.44E+11 | 6.87E+11 | 2.71E+11 | 6.55E+11 | 9.09E+10 | 7.65E+10 | 2.74E+10 |
|  | q4 | 3.09E+12 | 4.77E+11 | 1.63E+12 | 7.42E+11 | 6.89E+11 | 2.7E+11 | 6.44E+11 | 9.11E+10 | 7.6E+10 | 2.43E+10 |
| 2024 | q1 | 3.09E+12 | 4.76E+11 | 1.64E+12 | 7.38E+11 | 6.92E+11 | 2.69E+11 | 6.37E+11 | 9.06E+10 | 7.53E+10 | 2.11E+10 |
|  | q2 | 3.09E+12 | 4.74E+11 | 1.65E+12 | 7.33E+11 | 6.93E+11 | 2.69E+11 | 6.23E+11 | 8.97E+10 | 7.42E+10 | 1.77E+10 |
|  | q3 | 3.09E+12 | 4.73E+11 | 1.67E+12 | 7.28E+11 | 6.93E+11 | 2.68E+11 | 6.14E+11 | 8.85E+10 | 7.28E+10 | 1.45E+10 |
|  | q4 | 3.09E+12 | 4.72E+11 | 1.69E+12 | 7.24E+11 | 6.94E+11 | 2.68E+11 | 6.07E+11 | 8.71E+10 | 7.15E+10 | 1.17E+10 |
| 2025 | q1 | 3.09E+12 | 4.71E+11 | 1.71E+12 | 7.2E+11 | 6.93E+11 | 2.68E+11 | 6.01E+11 | 8.58E+10 | 7.05E+10 | 9.36E+09 |
|  | q2 | 3.09E+12 | 4.7E+11 | 1.72E+12 | 7.18E+11 | 6.92E+11 | 2.68E+11 | 5.98E+11 | 8.47E+10 | 6.97E+10 | 7.67E+09 |
|  | q3 | 3.1E+12 | 4.69E+11 | 1.72E+12 | 7.18E+11 | 6.9E+11 | 2.68E+11 | 5.99E+11 | 8.38E+10 | 6.92E+10 | 6.65E+09 |
|  | q4 | 3.1E+12 | 4.68E+11 | 1.72E+12 | 7.18E+11 | 6.88E+11 | 2.69E+11 | 6.03E+11 | 8.33E+10 | 6.88E+10 | 6.27E+09 |

According to the table 3 results for imports in OECD selected countries and Iran, the forecast imports trend in OECD selected countries and Iran is as follows over 20 seasons (from 2021 to 2025):

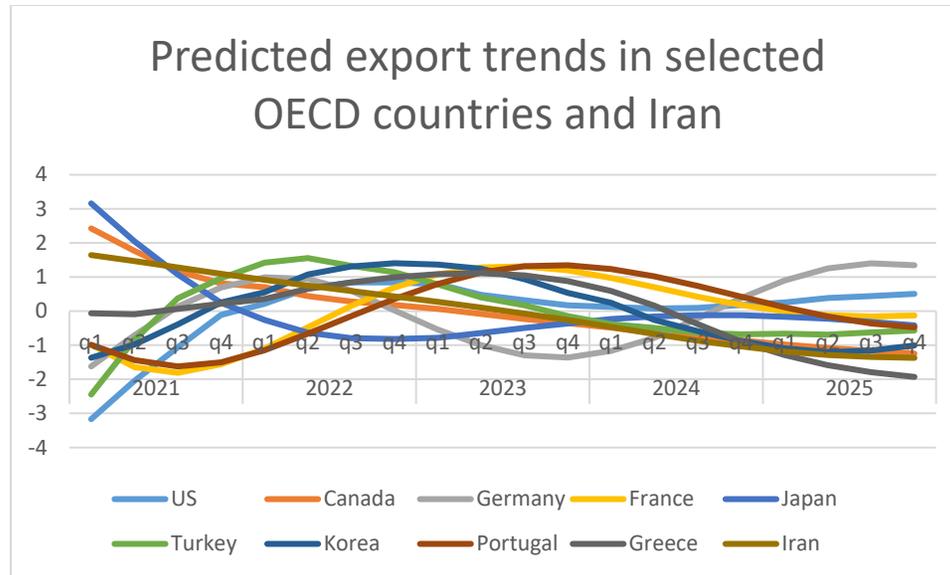

**Figure 2. Predicted export trends in selected OECD countries and Iran (**

Source: Author's Calculations)

According to the analyses, diagrams related to MSE and the forecast data including train data, test data and their regression for Iran are as follows. Considering the results, regression coefficient between target values and LSTM network output is 0.99 for all data.

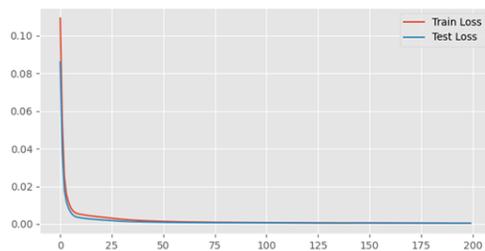 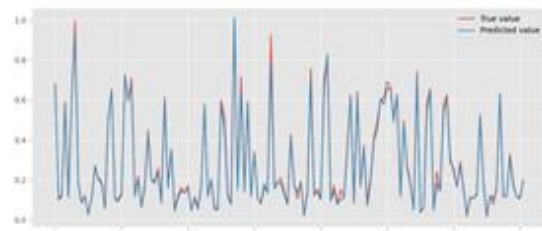 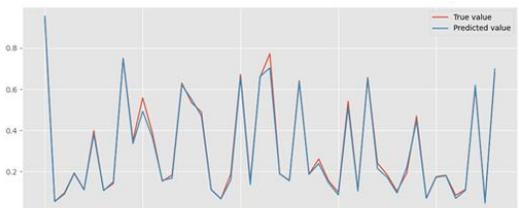

**Figure 3. MSE diagram for Test and Train Data Diagram    (Source: Author's Calculation)**

**Figure 4. Train Data Forecasting (Source: Author's Calculation)**

**Figure 5. Test Data Forecasting (Source: Author's Calculation)**

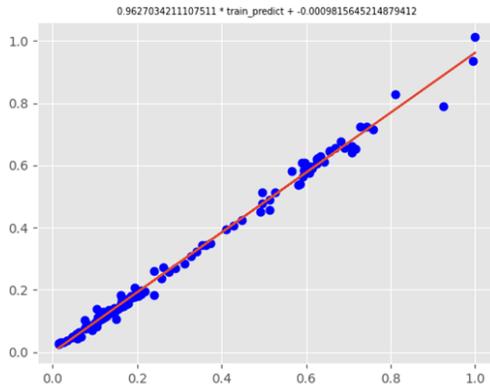

**Figure 6. Train Data Regression Diagram**

(Source: Author's Calculation)

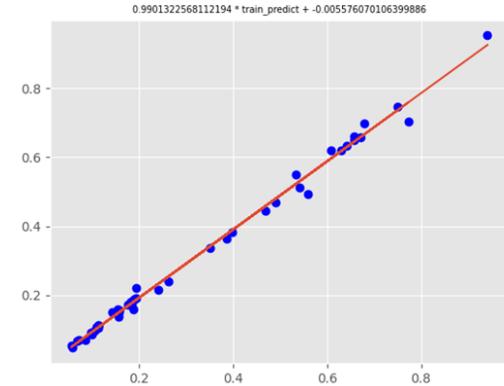

**Figure 7. Test Data Regression Diagram**

(Source: Author's Calculation)

6. **Conclusion**

Since the focus of economic and natural crises effect is revealed through consumption, so the imports is consequently influenced, therefore, it is important to investigate the changes trend of flows have been recently happened in the world on the imports of countries which are a combination of developed and developing countries.

Due to the structure of imports model, variables influential on it have nonlinear structure so this study has used LSTM neural network. Imports time series data collected from 1970 to 2019 in the selected countries and after implementation in the network, forecast accuracy obtained 99%. LSTM neural network has shown the best performance in correct forecasting rather than other traditional methods. Considering very suitable results of LSTM neural network on imports data with much volume, it is suggested to use this method which is able to encounter with data with time dependency (time series) and also cross-sectional effect at the same time with time dependency. The results taken from the output of this model are that with consumption reduction in crises flow, it is time-consuming to correct and improve it to be able to influence the imports, so imports flow is fluctuating in the years after Covid-19. Since imports is a good motive for government economy, it is possible to release recession has been created due to economic crises, so figure 2 shows that imports trend in most countries is ascending in 2022.

And since Import is a form of demand, it has a direct relationship with income, and during the years of the existence of Covid, it has a downward trend, and after that, import has an upward trend, which indicates that the economy will move towards more growth, more production and more income, as a result, Import will be increase.

7. **References**

Alam, T. 2019. Forecasting exports and imports through artificial neural network and autoregressive integrated moving average, *Decision Science Letters* 8. no. 3: 249-260.

**10.5267/j.dsl.2019.2.001**

Carrillo, M. R. 2020. Artificial intelligence: From ethics to law. *Telecommunications Policy* 44. no. 6: 101937.

Fan, D., H. Sun, J.Yao, K. Zhang, X. Yan, and Z. Sun. 2021. Well production forecasting based on ARIMA-LSTM model considering manual operations. *Energy* 220: 119708.

Ghaderzade, H. and Sh. Ghasiry darbande. 2021. Investigating the effect of oil price fluctuations on the variables of the agricultural sector by ARIMAX model and neural network in the period 1995-2015. *Agricultural economics research* 12. no. 4.

Greff, K., R. K. Srivastava, J. Koutník, B. R. Steunebrink, and J. Schmidhuber. 2016. LSTM: A search space odyssey. *IEEE transactions on neural networks and learning systems* 28. no.10: 2222-2232.

Haenlein, M. and A. Kaplan. 2021. Artificial intelligence and robotics: Shaking up the business world and society at large. *Journal of Business Research* 124: 405-407.

Heid, B. and I. Vozzo. 2020. The international trade effects of bilateral investment treaties. *Economics Letters* 196: 109569.

Koliadenko, P. 2021. Financial time series forecasting using hybrid ARIMA and Deep Learning models. *Master thesis*

Lot, R., F. Pellegrini, Y. Shaidu and E. Küçükbenli. 2020. Panna: Properties from artificial neural network architectures. *Computer Physics Communications* 256: 107402.

Luchko, M. R., N. Dziubanovska, and O. Arzamasova. 2021. Artificial Neural Networks in Export and Import Forecasting: An Analysis of Opportunities, *In 2021 11th IEEE International Conference on Intelligent Data Acquisition and Advanced Computing Systems: Technology and Applications (IDAACS)* 2: 916-923.

Moran, C. 1989. Imports under a foreign exchange constraint. *The World Bank Economic Review* 3. no. 2: 279-295.

Mustak, M., J. Salminen, L. Plé, and J. Wirtz. 2021. Artificial intelligence in marketing: Topic modeling. *scientometric analysis, and research agenda. Journal of Business Research*124: 389-404.

Nikaein, M. 2017. Identifying and prioritizing factors affecting importers' decision to import clothes to the Iranian market. *Master thesis*, Shahid Beheshti University.

Nikitin, N. O., P.Vychuzhanin, M. Sarafanov, I. S. Polonskaia, I. Revin, I. V Barabanova, G. Maximov, A.V. Kalyuzhnaya, and A. Boukhanovsky. 2022. Automated evolutionary approach for the design of composite machine learning pipelines. *Future Generation Computer Systems* 127: 109-125.

Paschen, J., M. Wilson, and J. J. Ferreira. 2020. Collaborative intelligence: How human and artificial intelligence create value along the B2B sales funnel. *Business Horizons* 63. no. 3: 403-414.

Prior, D. D., and J. Keränen. 2020. Revisiting contemporary issues in B2B marketing: It's not just about artificial intelligence. *Australasian Marketing Journal (AMJ)* 28. no. 2: 83-89.

Redmond, T. and M. A. Nasir. 2020. Role of natural resource abundance, international trade and financial development in the economic development of selected countries. *Resources Policy* 66: 101591.

Shafeei, B. Y. Bostan, A. Fattahi Ardakani, D. Jahangir Pour and R., Erfani Moghaddam. 2020. Forecasting and investigating the effect of real exchange rate uncertainty on the import of Iran's agricultural sector. *Agricultural Economic Research Quarterly* 12. no. 3: 125-150.

Shen, M. L., C. F. Lee, H. H. Liu, P. Y. Chang, and C. H. Yang. 2021. Effective multinational trade forecasting using LSTM recurrent neural network. *Expert Systems with Applications* 182: 115199.

Siami-Namini, S., N. Tavakoli, and A. S. Namin. 2019. The performance of LSTM and BiLSTM in forecasting time series. *In 2019 IEEE International Conference on Big Data (Big Data)*.3285-3292.

Spanlo, H. and A. Ghanbari. 2011. Investigating the factors affecting Iran's import demand by separating intermediate. *capital and consumer goods, Business Journal* 15. no 57: 209-233.

Stalidis, G., D. Karapistolis, and A. Vafeiadis. 2015. Marketing decision support using Artificial Intelligence and Knowledge Modeling: application to tourist destination management. *Procedia-Social and Behavioral Sciences* 175: 106-113.

Tsai, C. W., Y. P. Chen, T. C. Tang, and Y. C. Luo. 2021. An efficient parallel machine learning-based blockchain framework. *ICT Express* 7. no. 3: 300-307.

Urrutia, J. D., A. M. Abdul, A. M. and J. B. E. Atienza. 2019. Forecasting Philippines imports and exports using Bayesian artificial neural network and autoregressive integrated moving average. *In AIP Conference Proceedings* 2192. no. 1: 090015.

Vogl, M., P. G. Rötzel, and S. Homes. 2022. Forecasting performance of wavelet neural networks and other neural network topologies: A comparative study based on financial market data sets. *Machine Learning with Applications* 8: 100302.

Zhang, D. and S. Lou. 2021. The application research of neural network and BP algorithm in stock price pattern classification and prediction. Future Generation Computer Systems 115: 872-879.

Zinatian, Gazal. 2015. Analysis of the relationship between Iran's maritime transport, economic growth and international trade. *Master thesis*, Khorramshahr University of Marine Sciences and Techniques.

Hoang, A. T., Nižetić, S., Ong, H. C., Tarelko, W., Le, T. H., Chau, M. Q., & Nguyen, X. P. (2021). A review on application of artificial neural network (ANN) for performance and emission characteristics of diesel engine fueled with biodiesel-based fuels. *Sustainable Energy Technologies and Assessments*, *47*, 101416.